\documentclass[oupdraft]{bio}
\usepackage{url}
\usepackage{hyperref} 
\usepackage{float}
\usepackage{xcolor}
\usepackage{subcaption}
\history{}

\begin{document}
\makeatletter
\def\ps@plain{%
  \def\@oddhead{}%
  \def\@evenhead{}%
  \def\@oddfoot{\gridpl\copyl}%
  \def\@evenfoot{}%
}

\makeatletter
\def\copyl{}
\makeatother

\makeatother
\title{Enhancing comorbidity network inference with risk-enriched health trajectories embedding}

\author{Nicole Fontana$\ast$$^{1,2}$, Alessia Mapelli$\ast$$^{1,2}$, Emanuele Di Angelantonio$^{2,3,4,5,6}$, Francesca Ieva$^{1,2}$\\[4pt]
\textit{$^1$MOX, Department of Mathematics, Politecnico di Milano, Milan, Italy}\\
\textit{$^2$Health Data Science Research Centre, Human Technopole, Milan, Italy}\\
\textit{$^3$British Heart Foundation Cardiovascular Epidemiology Unit, Department of Public Health and Primary Care, University of Cambridge, Cambridge, UK} \\
\textit{$^4$Victor Phillip Dahdaleh Heart and Lung Research Institute, University of Cambridge, Cambridge, UK}\\
\textit{$^5$British Heart Foundation Centre of Research Excellence, University of Cambridge, Cambridge, UK}\\
\textit{$^6$National Institute for Health and Care Research Blood and Transplant Research Unit in Donor Health and Behaviour, University of Cambridge, Cambridge, UK}
\\[2pt]
{alessia.mapelli@polimi.it}\\
\textit{$^\ast$These authors contributed equally to this work.}}

\markboth%
{N. Fontana and A. Mapelli}
{Enhancing comorbidity network inference }

\maketitle


\begin{abstract}
{Multimorbidity poses a growing challenge for individual health, reducing quality of life and increasing treatment burden, resulting in a multiplicative impact on healthcare system management and fragmented care trajectories. Comorbidity networks could provide crucial insight into characterising multimorbidity and disease relationships. However, existing approaches to comorbidity network construction face critical limitations: they overlook temporal information by relying on cross-sectional statistics, produce biased association estimates by ignoring confounding due to shared risk factors, and fail to distinguish between direct and indirect disease associations, thereby yielding fully connected networks. To address these limitations, we develop a methodological framework for population-level disease network inference that uses individual health trajectories to learn disease associations, capturing semantic similarity and temporal co-occurrence. 
Sparse network estimation is achieved via Gaussian Graphical Models with Lasso regularisation, informed by prior clinical knowledge on shared risk factors derived from a dedicated confounding evaluation step. Applied to UK Biobank data comprising 24 cardiometabolic diseases and 76 risk factors, the resulting network revealed clinically meaningful disease patterns. Topological analysis identifies key pathological hubs, reveals potential actionable targets for multimorbidity management, and identifies four distinct disease communities that align with the established cardiometabolic taxonomy. Building on this community structure, we derive community-based patient representations that capture disease progression dynamics. Clustering these representations reveals four progression phenotypes with significantly different long-term survival trajectories, highlighting the potential of the framework for risk stratification and personalised care.
}
\end{abstract}

{Cardiometabolic diseases; Comorbidity network; Gaussian Graphical Models; Longitudinal data; Prior Clinical knowledge; Word2Vec embeddings}

\section{Introduction}\label{sec:introduction}
Multimorbidity, defined as the co-occurrence of two or more clinical conditions, affects approximately 25\% of adults and over 65\% of those aged 65 years or older~\citep{barnett2012epidemiology}. Cardiovascular and cardiometabolic diseases represent a particularly significant burden, accounting for nearly 18 million deaths annually worldwide and frequently presenting alongside metabolic, renal, and systemic disorders~\citep{roth2020global}. Understanding how these conditions develop and interact over time, and how underlying risk factors shape these patterns, is essential for developing targeted prevention strategies, optimising treatment protocols, and improving patient outcomes in complex clinical scenarios~\citep{PradosTorres2014, Yan2022}. Addressing these challenges requires statistical methods capable of extracting structured patterns of disease relationships from large-scale longitudinal patient data. 

The study of comorbidity networks has gained substantial attention as a powerful framework for characterising multimorbidity and disease relationships~\citep{DelValle2019}. Networks represent diseases as nodes and their associations as edges, providing a systems-level view of multimorbidity patterns~\citep{Barabasi2011}. Traditional approaches to comorbidity network construction typically rely on cross-sectional prevalence or incidence data, in which disease associations are computed as simultaneous presence rates within the study cohort at a specific time point~\citep{hidalgo2009dynamic, goh2007human}. Recent methodological advances have incorporated electronic health records and administrative databases to construct large-scale disease networks, leveraging statistical measures such as relative risk, odds ratios, phi-correlation, or Pearson correlation to quantify disease relationships~\citep{jensen2014temporal, siggaard2020disease, VargasFernandez2026Network, Burke2023}. While these methods have advanced our understanding of disease co-occurrence and revealed meaningful comorbidity structures, they face three critical limitations. First, most existing methods do not account for the timing and order of events across patients' lifetimes, relying on aggregate lifelong prevalence statistics or single-time-point measurements, such as simple pairwise co-occurrence counts~\citep{jeong2017network, dervic2025comorbidity}. Ignoring the temporal progression patterns and semantic disease relationships results in incomplete characterisations that fail to reflect the dynamic nature of multimorbidity development. Second, disease associations estimated solely from disease-based summary statistics measures are vulnerable to confounding by shared risk factors~\citep{vanderweele2013definition, pearl2009causality}. For example, two diseases may frequently co-occur not because of direct pathophysiological connections, but because they share common antecedents, such as lifestyle factors or other comorbid conditions. Traditional network construction methods that ignore this confounding produce biased estimates, artificially inflating the association between diseases with shared risk profiles. Third, typical estimation methods evaluate pairwise association, producing fully connected networks that capture both direct and indirect associations between diseases, making it difficult to distinguish genuine relationships from correlations induced by other conditions in the network. A common approach to address this issue is to apply arbitrary global thresholding for network sparsification~\citep{masuda2025}; however, the resulting network structure depends heavily on the threshold, potentially removing clinically relevant connections while retaining indirect or spurious associations.

To address these limitations, we develop a comprehensive pipeline for disease network inference from longitudinal patient data, integrating three methodological innovations. First, we leverage complete disease trajectories, processed as sequences, as input to capture the chronological progression of disease manifestations over the life course. We adapt Word2Vec embeddings, originally developed for natural language processing~\citep{mikolov2013efficient}, to longitudinal clinical sequences, fine-tuning pretrained biomedical embeddings~\citep{zhang2019biowordvec} to learn dense vector representations that jointly encode temporal co-occurrence patterns and semantic disease relationships. Second, we introduce a principled framework for confounding identification. By quantifying the contribution of shared risk factors to disease-pair associations, we derive interpretable disease-specific confounder sets, which we term prior clinical knowledge, and use them to explicitly adjust network estimation. 
Third, we employ Gaussian Graphical Models (GGMs) with Lasso regularization~\citep{meinshausen2006high, mapelli2026prior} to estimate a sparse network capturing only direct conditional dependencies. Unlike arbitrary thresholding approaches, this framework provides a statistically principled mechanism for network sparsification. By integrating the estimated prior clinical knowledge into a neighbourhood selection framework, the resulting network more accurately captures direct pathophysiological relationships after adjustment for shared risk factor burden. 


We apply this framework to construct a risk-adjusted population-level cardiometabolic disease network using longitudinal data from the UK Biobank, incorporating 24 primary cardiometabolic diseases and 76 risk factors. We then perform topological analyses on the estimated network, applying node centrality measures to assess the systemic importance of individual diseases and community detection to uncover distinct pathophysiological domains. Building on the resulting community structure, we derive a synthesised individual-level representations that capture how patients transition between disease domains over time. Finally, we cluster these community-driven fingerprints to characterise typical disease progression patterns and evaluate their prognostic value for long-term survival.

The remainder of this paper is structured as follows. Section~\ref{sec:data} describes the UK Biobank cohort, phenotype definitions, and data preprocessing. Section~\ref{sec:method} presents our methodological framework, including sequence construction, word embedding strategy, cosine-similarity-based confounder identification, GGM-based network estimation with prior clinical knowledge, and topological and downstream analysis. Section~\ref{sec:results} reports results for the risk-adjusted cardiometabolic disease networks, including confounding identification, network topology, community structure, and trajectory-based patient clustering and survival analysis. Section~\ref{sec:discussion} discusses methodological implications, limitations, and future directions.


\section{Data and phenotype definition}\label{sec:data}
The UK Biobank is a large-scale prospective cohort study comprising approximately 500,000 participants aged 40-69 years at recruitment between 2006 and 2010 across the United Kingdom~\citep{Sudlow2015UKBiobank}. This population-based resource is among the most comprehensive biomedical databases worldwide, systematically integrating demographic information, lifestyle questionnaires, physical measurements, biological samples, multi-omics data, and longitudinal electronic health record (EHR) linkage. We defined 24 primary cardiometabolic diseases as the core focus of the comorbidity network, alongside 76 risk factors selected based on established epidemiological literature. Phenotype definitions were developed through expert clinical consultation to ensure consistency with diagnostic criteria and epidemiological standards. We adopted a multi-source ascertainment strategy that integrates three distinct clinical domains: ICD-10 diagnosis codes and OPCS-4 procedure codes indicating surgical or interventional treatment (1) from hospital inpatient records; (2) from primary care data; and (3) laboratory measurements. 
This integrated phenotypic representation harmonises information across multiple data modalities and longitudinal sources, thereby improving disease ascertainment sensitivity relative to hospital-only records. The 24 primary diseases span six clinical categories: major cardiovascular conditions ($n=6$), valvular heart diseases ($n=3$), vascular diseases ($n=5$), advanced cardiovascular interventions ($n=2$), metabolic conditions ($n=2$, plus hypertension), and progressive stages of renal dysfunction ($n=5$).
In addition to the primary diseases, 76 risk factors were considered to adjust for confounding, including demographic and socioeconomic factors, family history of cardiovascular diseases, lifestyle and anthropometric factors, clinical biomarkers, and comorbidities. Complete definitions and life-course incidence data for each disease are provided in Supplementary Table~\ref{ST:table1} for the primary diseases and in Supplementary Table~\ref{ST:table2} for risk factors. Longitudinal incidence distribution plots for each primary disease are shown in Supplementary Figure~\ref{supp_mat:SF0}. All phenotypes were mapped to the Medical Subject Headings (MeSH) vocabulary to enable integration with BioWordVec pretrained embeddings. After MeSH mapping, the final vocabulary comprised 24 unique main diseases and 98 unique events linked to risk factors levels. Information about the MeSH mapping is provided in Supplementary Tables~\ref{ST:table1} and~\ref{ST:table2}. We recorded 544,716 primary disease events and 3,285,210 risk factor events. Starting from 502,297 UK Biobank participants, we excluded individuals with fewer than three recorded main disease events during the life course. The final cohort comprised 113,973 participants. For each participant, we constructed longitudinal event sequences capturing temporal progression throughout their clinical history; details are provided in Section~\ref{subsection:seq_con}. 

\section{Methods}\label{sec:method}
\begin{figure}[t]
    \centering
    \hspace*{-1.8cm}\includegraphics[width=1.3\linewidth]{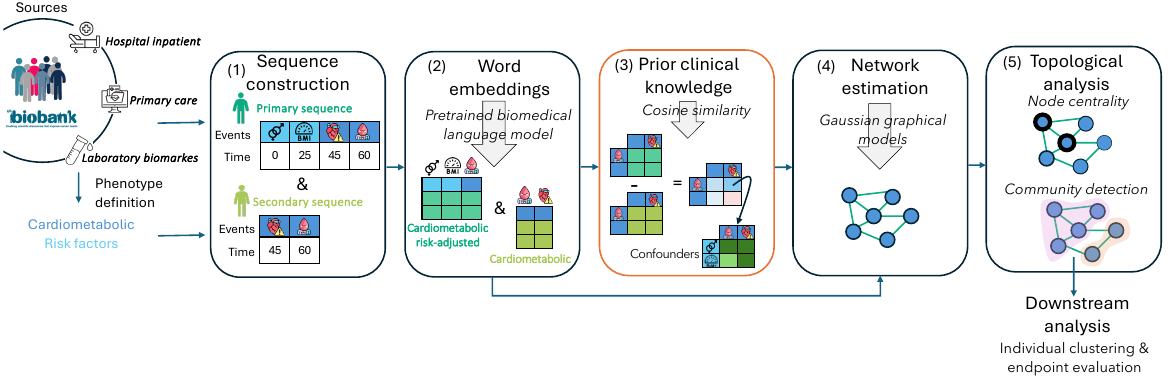}
    \caption{Schematic representation of the methodological framework for risk-adjusted comorbidity network inference. The workflow proceeds through five stages after the definition of phenotypes, integrating multimodal data sources: (1) Primary and secondary sequence construction. (2) Embedding extraction of clinical events by fine-tuning pretrained biomedical language models on sequences. (3) Prior clinical knowledge extracted from the association of risk factors with the similarity difference between risk-adjusted and disease-only embedding matrices. (4) Network estimation via Gaussian Graphical Models with prior clinical knowledge. (5) Topological analysis via node centrality, community detection, and downstream analysis.}
    \label{fig:pipeline}
\end{figure}

In this section, we present the methodological pipeline for estimating disease comorbidity networks from individual-level longitudinal clinical data while adjusting for confounding due to shared risk factors. As shown in Figure~\ref{fig:pipeline}, it consists of five main stages, each addressing a key challenge in translating longitudinal patient data into interpretable disease networks. First, we construct temporally ordered event sequences from longitudinal clinical data (Section~\ref{subsection:seq_con}). Second, we embed sequences of disease and risk factor events into low-dimensional latent representations (Section~\ref{subsection:word_embedding}).
Third, we leverage cosine similarity between learned embeddings to identify, validate, and select disease-specific confounders, to construct prior clinical knowledge (Section~\ref{subsection:prior_knowledge}). Fourth, we estimate a sparse network of conditional dependencies via a Gaussian Graphical Model (GGM), explicitly incorporating prior clinical knowledge (Section~\ref{subsec:sparsification}). Finally, we characterise the modular architecture of the inferred network through topological analysis and derive patient-level representations from community structure to support individual-level clustering and survival prediction (Section~\ref{subsec:topology}).

\subsection{Sequence construction}\label{subsection:seq_con}
For each patient $p \in \{1, \ldots, P\}$, we constructed a temporal event sequence (Figure~\ref{fig:pipeline}, Step 1):
\begin{equation}
S_p = \{(t_1, v_1), (t_2, v_2), \ldots, (t_{n_p}, v_{n_p})\},
\end{equation}
where $n_p \in \mathbb{N}$ denotes the number of recorded events, $v_j \in \mathcal{V}$ represents the event identifier (disease or risk factor), and $t_j \in \mathbb{R}_+$ denotes time from birth in days. Sequences are ordered chronologically such that $t_1 \le t_2 \le \cdots \le t_{n_p}$. This temporal ordering enables the embedding model to capture both co-occurrence patterns and disease progression dynamics. We constructed a primary sequence $S_p^{(\mathcal{D}+\mathcal{R})}$ containing events from the full vocabulary $\mathcal{V} = \mathcal{D} \cup \mathcal{R}$, where $\mathcal{D}$ denotes the set of cardiometabolic diseases and $\mathcal{R}$ the set of risk factors, yielding a vocabulary size $|\mathcal{V}^{(\mathcal{D}+\mathcal{R})}| = 122$. To evaluate the effect of risk factor adjustment, we also constructed a secondary sequence $S_p^{(\mathcal{D})}$, restricted to cardiometabolic diseases only, with vocabulary size $|\mathcal{V}^{(\mathcal{D})}| = 24$. In the following, we denote events $v_j \in \mathcal{D}$ by $d_j$ and events $v_j \in \mathcal{R}$ by $r_j$.

\subsection{Word embeddings}\label{subsection:word_embedding}
In natural language processing, representation learning maps words into a continuous, low-dimensional embedding vector space where geometric proximity reflects semantic similarity derived from contextual co-occurrence~\citep{liu2023representation}. We adapted the Word2Vec skip-gram architecture to event sequences by conceptualising medical events as ``words'' and entire individual-level sequences as ``sentences'' (Figure~\ref{fig:pipeline}, Step 2). In this formulation, each event's training context is defined by a fixed window of $c$ neighbouring events preceding and following the considered one~\citep{bornet2025comparing}. The choice of this framework is motivated by Word2Vec's ability to generate embeddings, in which entities appearing in similar clinical contexts, such as conditions with shared pathophysiological precursors, are mapped closely in the embedding space. This approach captures temporal relationships between diseases directly from raw event sequences, without requiring more complex architectures such as recurrent or transformer-based models~\citep{mienye2024recurrent, madan2024transformer}, thereby balancing representational power with computational efficiency.
We leveraged the pretrained BioWordVec model~\citep{zhang2019biowordvec}, which trains embeddings combining 27 million PubMed abstracts with the Medical Subject Headings (MeSH) structure, thereby providing domain-specific biomedical knowledge. Individual-level sequences are tokenised into constituent events, and multi-word events are further tokenised into constituent words (e.g., ``Chronic Kidney Disease'' $\rightarrow$ [``Chronic'', ``Kidney'', ``Disease'']). To produce an embedding that captures both domain-specific semantics and temporal relationships, individual tokenised sequences are exploited to fine-tune the BioWordVec model using the original skip-gram configuration with embedding dimension $d=200$, context window size $c=5$, and standard learning rate schedule. To balance cohort-specific adaptation with preservation of biomedical semantics, we monitored the cosine similarity between pretrained and fine-tuned embeddings:
\begin{equation}
\text{sim}_{\text{preserved}}^{(t)} = \frac{1}{|\mathcal{V}|} \sum_{v \in \mathcal{V}} \frac{\mathbf{e}_v^{(t)} \cdot \mathbf{e}_v^{(0)}}{\|\mathbf{e}_v^{(t)}\| \|\mathbf{e}_v^{(0)}\|},
\end{equation}
where $\mathbf{e}_v^{(t)}$ denotes the embedding at epoch $t$ and $\mathbf{e}_v^{(0)}$ the pretrained embedding. Fine-tuning is stopped when $\text{sim}_{\text{preserved}}^{(t)} < 0.70$, ensuring at least $70\%$ semantic retention, satisfied after 4 epochs in our specific use case. For phenotypes composed of multiple words (e.g., ``Chronic Kidney Disease"), the phenotype embedding is computed as the average of all the constituent words embeddings (e.g., $\{\text{``Chronic"; ``Kidney"; ``Diseases"}\}$).
The method is independently applied to the primary and secondary sequences to extract two embedding spaces: the first incorporating risk factor information, the second focused solely on cardiometabolic diseases.

\subsection{Prior clinical knowledge via confounding detection}\label{subsection:prior_knowledge}
Starting from the two embedding spaces, in this step, we construct what we term prior clinical knowledge, identifying shared risk factors that may confound disease associations (Figure~\ref{fig:pipeline}, Step 3).

\subsubsection{Differential similarity matrix construction}\label{subsub:network_con}
Following established practices in the literature~\citep{gundougan2020research}, we computed pairwise cosine similarity between phenotype embeddings $\{\mathbf{e}_1, \ldots, \mathbf{e}_D\}$ to construct a similarity matrix $A \in [0,1]^{D \times D}$: 
\begin{equation}\label{eq:cosine_similarity}
A_{ij} = \text{sim}(\mathbf{e}_i, \mathbf{e}_j) = \frac{\mathbf{e}_i \cdot \mathbf{e}_j}{\|\mathbf{e}_i\| \|\mathbf{e}_j\|},
\end{equation}
whose entries reflect the overall semantic and contextual relatedness between diseases. Cosine similarity ranges from $-1$ (opposite identical) to $+1$ (identical), with values near 0 indicating orthogonality in the embedding space (no shared meaning). We constructed two disease similarity matrices: $A^{(\mathcal{D}+\mathcal{R})}$, derived from embeddings learned from the primary sequences containing both cardiometabolic diseases and risk factors, and $A^{(\mathcal{D})}$, derived from embeddings trained on the secondary disease-only sequences.


\subsubsection{Conceptual framework for confounding definition}\label{subsub:conf_framework}
We hypothesize that confounding effects in the embedding space manifest as follows: if diseases $d_i, d_j \in \mathcal{D}$ are associated due to a shared common risk factor $r_k \in \mathcal{R}$, then similarity computed from disease-only sequences, $A_{ij}^{(\mathcal{D})}$, is expected to be inflated related to the ``true" pathophysiological relationship. Jointly embedding diseases and risk factors redistributes part of this shared signal to the risk factor dimensions, yielding adjusted disease embeddings and corrected similarity estimates $A_{ij}^{(\mathcal{D}+\mathcal{R})}$.

\subsubsection{Statistical evaluation of the differential matrix}\label{subsub:stat_eval}
We quantified similarity changes as:
\begin{equation}\label{eq:similarity_difference}
\Delta_{ij} = A_{ij}^{(\mathcal{D}+\mathcal{R})} - A_{ij}^{(\mathcal{D})}.
\end{equation}
Negative values indicate confounding removal (similarity decreased after adjustment), while positive values suggest unmasking of true relationships previously obscured by noise. We assessed the global impact of risk adjustment using the Mean Absolute Difference (MAD) to quantify the average magnitude of change across all disease pairs and the Mean Directional Change (MDC) to assess systematic bias in adjustment direction: 
\begin{equation}
\text{MAD} = \frac{1}{D(D-1)/2} \sum_{i < j} |\Delta_{ij}|,\text{ } \text{ } \text{ } \text{MDC} = \frac{1}{D(D-1)/2} \sum_{i < j} \Delta_{ij},
\end{equation}
where negative MDC indicates systematic similarity decreases. We employed a Wilcoxon signed-rank test to provide formal statistical evidence of systematic change. To evaluate preservation of network structure, we computed the Spearman rank correlation between $A^{\mathcal{D}}$ and $A^{\mathcal{D+R}}$, where a high correlation indicates structural preservation despite local adjustments.

\subsubsection{Confounding detection and validation}\label{subsub:conf_det}
To validate that observed changes in similarity reflect genuine confounding removal rather than random noise, we implemented an analysis linking changes in disease similarity to the burden of shared risk factors. Disease pairs were first ranked according to $|\Delta_{ij}|$, and pairs within the 95th percentile of the distribution were considered highly affected by the adjustment procedure. Then, for each disease $d_i \in \mathcal{D}$, we identified the subset of associated risk factors $\mathcal{R}_i \subset \mathcal{R}$ based on embedding similarity. Specifically, a risk factor $r \in \mathcal{R}$ was considered associated with disease $d_i$ when $\text{sim}(\mathbf{e}_{d_i}, \mathbf{e}_{r}) > \tau$. In our analysis, $\tau$ was chosen as the median pairwise similarity among all risk factors. This thresholding strategy ensured that only disease-risk factor associations that were strong relative to the global correlation structure of the risk factor embedding space were retained for further analysis. For each disease pair $(d_i, d_j)$, we then quantified the proportion of shared risk factors as:
\begin{equation}\label{eq:shared_rf}
\text{SharedRF}_{ij} = \frac{|\mathcal{R}_i \cap \mathcal{R}_j|}{\min(|\mathcal{R}_i|, |\mathcal{R}_j|)}.
\end{equation}
Under our confounding framework, disease pairs with large changes in similarity are expected to share a greater burden of common risk factors. To formally evaluate this hypothesis, we used a Mann-Whitney U test assessing whether the distribution of $\text{SharedRF}_{ij}$ was stochastically different among highly affected pairs compared with all remaining pairs. A significant result would provide statistical evidence that the magnitude of the similarity correction is associated with shared risk factor burden, supporting the validity of the adjustment procedure.
Finally, to evaluate the robustness of the confounding detection threshold, the analysis was repeated across $\tau \in \{0.6, 0.7, 0.8\}$. 

\subsubsection{Confounder selection as prior clinical knowledge}\label{subsub:conf_selection}
Following the procedure described above, we retrieved, for each disease pair $(d_i, d_j)$, the set of identified confounders $\mathcal{C}_{ij} = \mathcal{R}_i \cap \mathcal{R}_j$. These pair-specific confounder sets are then aggregated at the disease level. For each disease $d_i \in \mathcal{D}$, we defined its overall confounder set as the union of all pairwise confounder sets involving $d_i$: $ \mathcal{C}_i^* = \bigcup_{j \in \mathcal{D} \setminus \{i\}} \mathcal{C}_{ij}$. The sets $\{\mathcal{C}_i^*\}_{i \in \mathcal{D}}$ constitute the prior clinical knowledge on confounding structure and can be directly incorporated into the network estimation procedure described in Section~\ref{subsec:sparsification}.

\subsection{Network estimation using Gaussian Graphical Models}\label{subsec:sparsification}
To estimate a sparse network of direct conditional dependencies within $\mathcal{D}$, we adopted a Gaussian Graphical Model (GGM) framework with Lasso (Figure~\ref{fig:pipeline}, Step 4). This method estimates partial correlations that quantify the association between two diseases after conditioning on all other variables, providing a principled mechanism for sparsification that effectively prunes edges that do not represent direct conditional dependencies. Treating each embedding dimension as an independent observation~\citep{zhelezniak2019correlation}, the embedding vectors $\{\mathbf{e}_v\}_{v \in \mathcal{V}}$ are assumed to follow a multivariate Gaussian distribution $ \mathbf{e} \sim \mathcal{N}(\boldsymbol{\mu}, \boldsymbol{\Sigma})$ where $\boldsymbol{\Sigma} \in \mathbb{R}^{(D+R) \times (D+R)}$ is the covariance matrix. Under this hypothesis, the conditional independence of the diseases is encoded in the reduced precision matrix 
$\boldsymbol{\Omega} = (\boldsymbol{\Sigma_{D})}^{-1}$ with $\boldsymbol{\Omega}_{ij} = 0$ if and only if embeddings $i$ and $j$ are conditionally independent given all other events. The multivariate normality assumption is usually evaluated using Shapiro-Wilk tests on each embedding dimension, with a significance threshold $\alpha = 0.05$. We employed a neighbourhood selection approach~\citep{mapelli2026prior} to estimate $\boldsymbol{\Omega}$ through $D$ independent Lasso regression models. In this stage, the predictor set for each disease regression could be informed by the prior clinical knowledge derived in Section~\ref{subsection:prior_knowledge}. For each phenotype $i \in \mathcal{D}$, we solved:
\begin{equation}\label{eq:neighborhood_selection}
\hat{\boldsymbol{\beta}}^{(i)} = \mathop{\arg \min}_{\boldsymbol{\beta} \in \mathbb{R}^{R+D-1}} \|\mathbf{e}_i - \mathbf{E}_{-i}\boldsymbol{\beta}\|_2^2 + \lambda \|\boldsymbol{\beta}\|_1,
\end{equation}
where $\mathbf{E}_{-i}$ is the matrix of predictors for disease $d_i$, comprising the embeddings of all other cardiometabolic diseases $\mathcal{D} \setminus \{d_i\}$ together with the embeddings of the disease-specific confounder set $\mathcal{C}_i^*$. Including $\mathcal{C}_i^*$ as covariates ensures that the inferred associations between $d_i$ and other diseases reflect direct relationships, after accounting for the shared risk factor burden identified in the prior knowledge step. The $\ell_1$-lasso penalty $\lambda > 0$ controls sparsity and can be selected via 10-fold cross-validation, minimising mean squared prediction error. The estimated neighbourhood of $d_i$ is defined using the OR rule to ensure symmetry: $\hat{E}_i = \{j : \hat{\beta}_j^{(i)} \neq 0 \text{ or } \hat{\beta}_i^{(j)} \neq 0\}$. The binary adjacency matrix describing the estimated network is then obtained as: $B_{ij} = \mathbb{I}(j \in \hat{E}_i)$.

\subsection{Network topology analysis and downstream analysis}\label{subsec:topology}
Once the population-level, risk-adjusted network is estimated, topological analysis is performed to characterise its structural properties. Building on its results, several downstream analyses can be performed; in our pipeline, we focus on patient subgroup identification task.

\subsubsection{Node centrality and communities detection}
To characterise the importance of individual diseases and uncover the modular architecture of the estimated network, we performed a topological analysis (Figure~\ref{fig:pipeline}, Step 5). 
Disease importance was quantified using two complementary centrality measures. Degree centrality ($C_D$) captures a node's local connectivity by counting its direct disease associations, whereas betweenness centrality ($C_B$) provides a global measure of influence by identifying diseases that act as bridges between distinct pathophysiological domains~\citep{freeman1978centrality}.
In addition, we identified cohesive disease modules by applying the fast greedy modularity optimisation algorithm~\citep{clauset2004finding}. This approach partitions the network into communities by maximising the modularity $Q$:
\begin{equation}
Q = \frac{1}{2m} \sum_{ij} \left(B_{ij} - \frac{C_D(i) C_D(j)}{2m}\right) \delta(c_i, c_j),
\end{equation}
where $m = \frac{1}{2}\sum_{ij} B_{ij}$ represents the total edge count, $c_i$ denotes the community assignment of node $i$, and $\delta(c_i, c_j) = 1$ if $c_i = c_j$ and 0 otherwise. The algorithm iteratively merges communities to maximise $Q$, producing a hierarchical dendrogram. We select the partition with maximum modularity as the final community structure.

\subsubsection{Trajectory-based patient clustering and survival}\label{subsub:transition}
Building on the detected community structure, we derived individual-level patient representations that capture disease progression dynamics between disease modules. Events within clinical sequences were mapped to their corresponding community.
Let $K$ denote the number of detected communities. For patient $p$, we constructed a $K \times K$ transition count matrix $T_p$, where entry $(i,j)$ contains the number of times a disease in community $i$ was immediately followed by a disease in community $j$ in that patient's event sequence. 
The matrix was then normalised by the total number of transitions observed for that patient, such that it reflects the relative frequency of each transition type within the patient's progression history, making matrices comparable across patients regardless of sequence length. The population-level transition structure was summarised in a $K\times K$ matrix of mean transitions, revealing which community-to-community movements are most common in the cohort and, in particular, whether patients tend to accumulate diseases within the same domain (high diagonal values) or systematically progress across domains (high off-diagonal values). Each patient's normalised transition matrix was flattened into a $K^2$-dimensional vector and used as input to $k$-means clustering. Patients assigned to the same cluster share similar disease progression dynamics. The optimal number of clusters was selected via the elbow criterion on the total within-cluster sum of squares. For each cluster, the mean transition matrix was computed to characterise the dominant progression pattern of each group. To assess whether disease progression dynamics captured by community transition patterns are associated with patient survival, we estimated Kaplan-Meier survival curves separately for each transition cluster. Patients who died were treated as events; surviving patients were censored at their individual administrative censoring date, defined as the time elapsed from birth to the loss of follow-up, computed specifically for each patient based on their year and month of birth.

\section{Results}\label{sec:results}
We applied our methodological framework to construct a cardiometabolic comorbidity network from UK Biobank longitudinal data. Staring from embeddings extracted from the fine-tuned BioWordVec model, as detailed in Section~\ref{subsection:word_embedding}, we compare risk-adjusted and disease-only cosine similarity matrices to validate that their differences reflect genuine confounding by shared risk factors and extract disease-specific confounder sets that constitute the prior clinical knowledge.


 
\begin{figure}[h]
    \centering
    \includegraphics[width=0.9\textwidth]{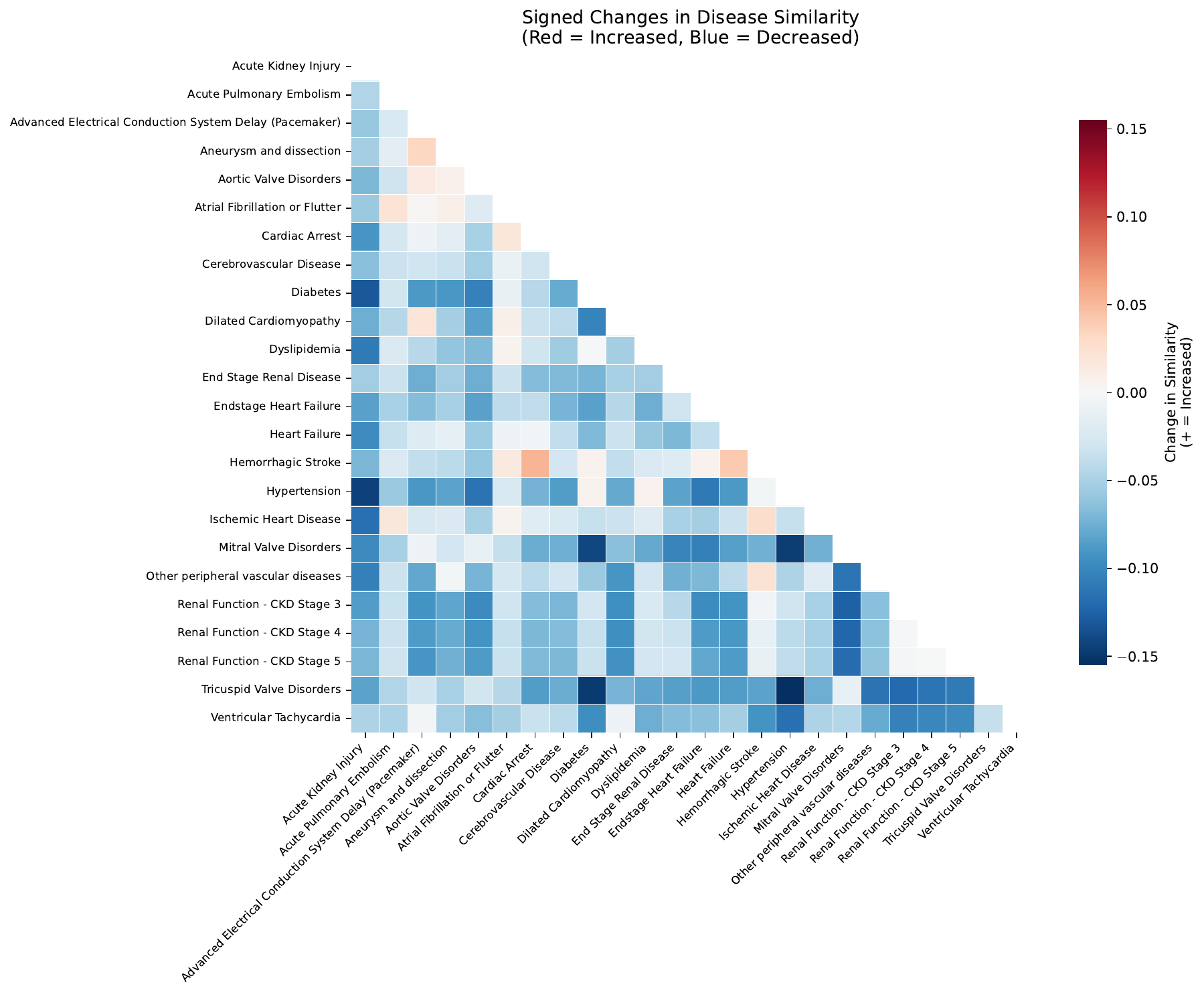}
    \caption{Signed changes in disease similarity after risk factor adjustment ($\Delta$). Blue indicates decreased similarity, red indicates increased similarity.}
    \label{fig:disease_similarity_changes_signed}
\end{figure}

The two disease similarity matrices used to identify confounders, $A^{(\mathcal{D+R})}$ derived from primary sequences and $A^{(\mathcal{D})}$ derived from disease-only sequences, are shown in Supplementary Figures~\ref{supp_mat:SF2} and~\ref{supp_mat:SF1}, respectively. 
The computed difference matrix $\Delta$ is shown in Figure~\ref{fig:disease_similarity_changes_signed} and the global network comparison metrics are summarised in Table~\ref{tab:matrix_level_stats}. In total, 252 of 276 disease pairs (91.3\%) exhibit decreased similarity after risk adjustment, and 24 pairs (8.7\%) show increased similarity. The MAD across all disease pairs is moderate, and the strongly negative MDC demonstrates a systematic decreasing tendency after risk adjustment. The Wilcoxon signed-rank test provides formal statistical evidence that these changes are non-random ($p = 2.20 \times 10^{-43}$), confirming our hypothesis that shared risk factors artificially inflate disease similarities in the unadjusted network. The Spearman rank correlation ($\rho_S = 0.77$) indicates that the fundamental ordering of disease relationships is largely preserved, suggesting that risk adjustment refines rather than fundamentally restructures the similarity matrix. The distribution of impact shows that 91.3\% of disease pairs experienced changes exceeding 1\%, with 9.4\% exceeding 10\%.

\begin{table}[h]
\centering
\begin{tabular}{lc}
\hline
\textbf{Metric} & \textbf{Value} \\
\hline
Mean Absolute Difference (MAD) & 0.056 \\
Mean Directional Change (MDC) & $-0.054$ \\
Wilcoxon signed-rank test & $p = 2.20 \times 10^{-43}$ \\
Spearman rank correlation & $\rho_S = 0.77$, $p = 9.68 \times 10^{-56}$ \\
\hline
\multicolumn{2}{l}{Impact distribution across disease pairs:} \\
\text{ } Pairs with change $> 1\%$ & 252 / 276 (91.3\%) \\
\text{ } Pairs with change $> 5\%$ & 150 / 276 (54.3\%) \\
\text{ } Pairs with change $> 10\%$ & 26 / 276 (9.4\%) \\
\hline
\end{tabular}
\caption{Matrix-level statistics comparing disease-only versus risk-adjusted similarity networks.}
\label{tab:matrix_level_stats}
\end{table}

Validating our confounding framework, we quantified shared risk factors for all disease pairs, setting $\tau = 0.73$. Table~\ref{tab:top_affected_pairs_shared} reports the highly affected disease pairs with the largest changes in similarity (top 5\%), with all these pairs showing decreased similarity. Pairs involving hypertension and diabetes mellitus dominate the ranking. The top-affected pairs share substantially more risk factors than lower-changed pairs (Mann-Whitney U test p-value = $2.57 \times 10^{-14}$), confirming that the magnitude of the similarity correction is associated with shared risk factor burden.

\begin{table}[h]
\centering
\small
\begin{tabular}{cccc}
\hline
\textbf{Rank} & \textbf{Disease pair} & \textbf{Change} & \textbf{SharedRF} \\
\hline
1 & Hypertension $\leftrightarrow$ Tricuspid Valve Insufficiency & -0.155 & 0.69 \\
2 & Diabetes Mellitus $\leftrightarrow$ Tricuspid Valve Insufficiency & -0.148 & 0.78 \\
3 & Mitral Valve Insufficiency $\leftrightarrow$ Hypertension & -0.147 & 0.78 \\
4 & Acute Kidney Injury $\leftrightarrow$ Hypertension & -0.144 & 0.79 \\
5 & Mitral Valve Insufficiency $\leftrightarrow$ Diabetes Mellitus & -0.141 & 0.84 \\
6 & Acute Kidney Injury $\leftrightarrow$ Diabetes Mellitus & -0.131 & 0.86 \\
7 & CKD Stage 3 $\leftrightarrow$ Mitral Valve Insufficiency & -0.127 & 0.96 \\
8 & CKD Stage 4 $\leftrightarrow$ Mitral Valve Insufficiency & -0.123 & 0.96 \\
9 & CKD Stage 3 $\leftrightarrow$ Tricuspid Valve Insufficiency & -0.120 & 0.97 \\
10 & Mitral Valve Insufficiency $\leftrightarrow$ CKD Stage 5 & -0.120 & 0.96 \\
11 & Coronary Disease $\leftrightarrow$ Acute Kidney Injury & -0.117 & 1.00 \\
12 & Hypertension $\leftrightarrow$ Ventricular Tachycardia & -0.117 & 0.68 \\
13 & Peripheral Vascular Diseases $\leftrightarrow$ Tricuspid Valve Insufficiency & -0.115 & 0.97 \\
14 & Hypertension $\leftrightarrow$ Aortic Valve Stenosis & -0.114 & 0.85 \\
\hline
\end{tabular}
\caption{Highly affected disease pairs (95th percentile), showing changes in similarity and SharedRF indicating the proportion of risk factors simultaneously associated with both diseases ($\tau = 0.73$).}
\label{tab:top_affected_pairs_shared}
\end{table}

To assess the robustness of our results, we performed a sensitivity analysis by varying the risk factor definition threshold, $\tau$, from $0.6$ to $0.8$. As shown in Supplementary Table~\ref{ST:shared_rf_sensitivity}: as $\tau$ increases, the number of risk factors associated with each disease and consequently the number of shared risk factors, decreases. Choosing the median for the threshold definition allows for a balance between including and excluding potential confounders. Notably, even as the number of individual risk factors per disease declines with higher $\tau$ values, the Mann-Whitney U test remains significant. This consistency demonstrates that our findings regarding confounding removal are robust to threshold specification. Finally, we extracted the disease-specific confounder sets, constituting the prior clinical knowledge shown in Supplementary Figure~\ref{supp_mat:SF3}.\\

\begin{figure}[htbp]
    \centering
    \begin{subfigure}[b]{0.49\textwidth}
        \centering
        \includegraphics[width=\textwidth]{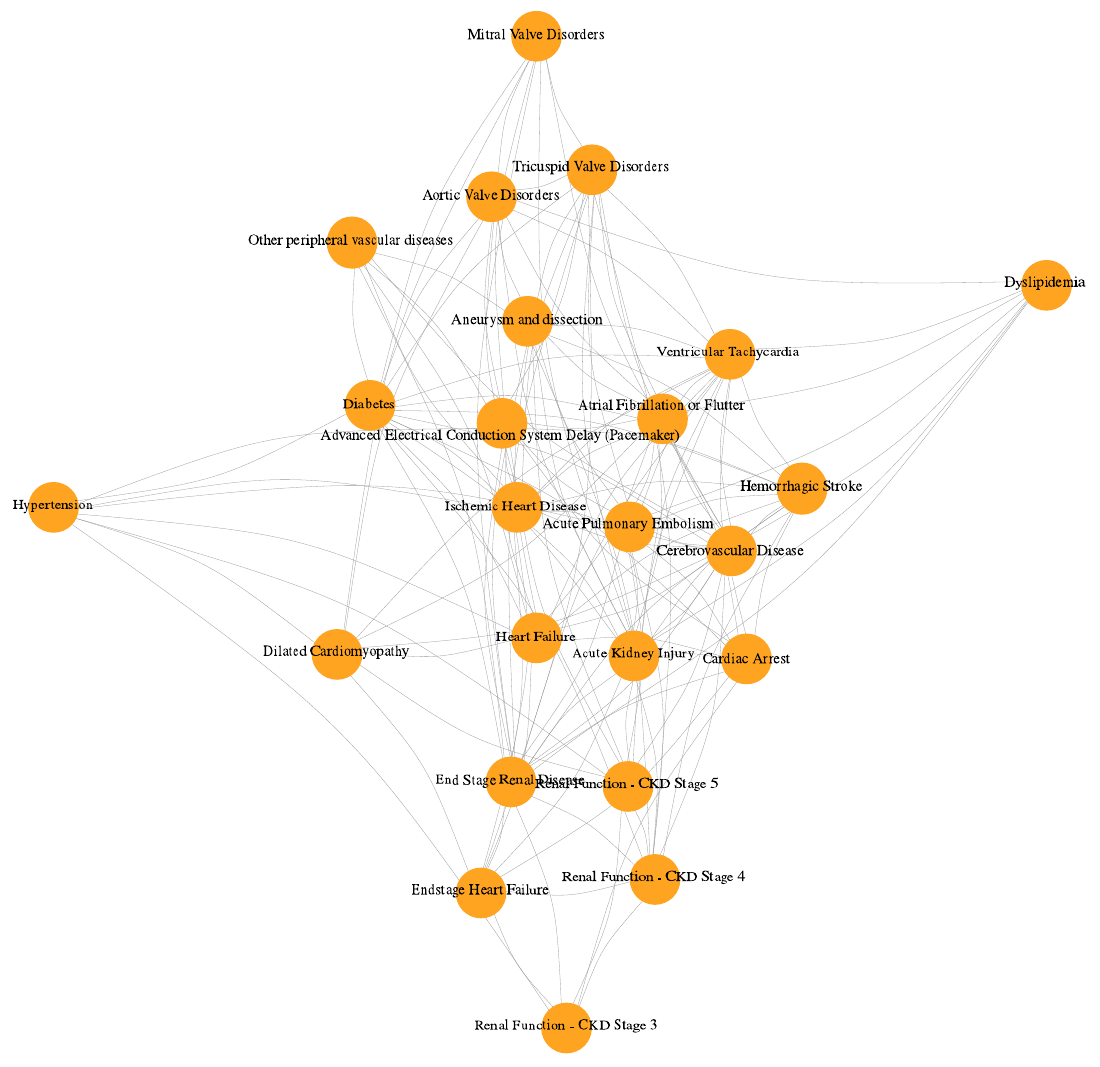}
        \caption{}
        \label{fig:GGM_comorb_network_heatmap1}
    \end{subfigure}
    \hfill
    \begin{subfigure}[b]{0.49\textwidth}
        \centering
        \includegraphics[width=\textwidth]{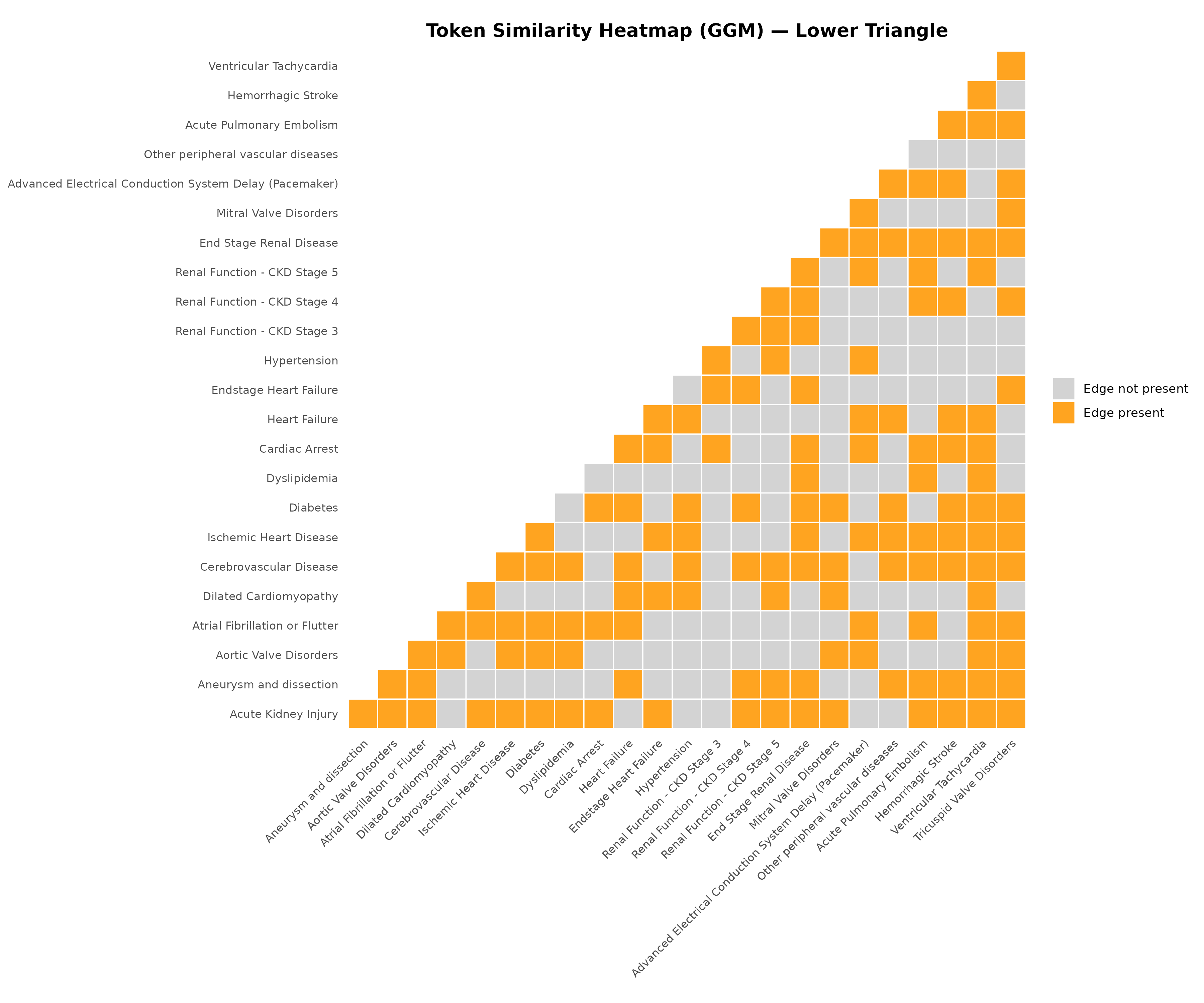}
        \caption{}
        \label{fig:GGM_comorb_network_heatmap2}
    \end{subfigure}
    \caption{
    Estimated cardiometabolic risk-adjusted disease network (a), and corresponding adjacency matrix (b).}
    \label{fig:combined_GGM}
\end{figure}

After assessing the normality of the extracted embeddings using Shapiro-Wilk tests, we estimated the sparse risk-adjusted cardiometabolic disease network using Gaussian Graphical Models with Lasso regularisation and incorporating prior clinical knowledge. The resulting network is shown in Figure~\ref{fig:combined_GGM}, along with the corresponding binary adjacency matrix. 
The resulting sparse network recovers clinically coherent disease results, consistent with established cardiometabolic taxonomy~\citep{roth2020global}: a renal disease progression pathway, a valvular disease module, and an ischemic-arrhythmic-heart failure triad.

To characterise the temporal ordering of disease progression within the estimated network, we computed a $24 \times 24$ disease-level transition matrix shown in Supplementary Figure~\ref{supp_mat:SF4}. Each cell represents the percentage of individuals whose trajectory included a transition from the source to the target disease. The resulting heatmap reveals clear asymmetries: several conditions, such as hypertension, diabetes, and ischemic heart disease,  show substantially higher outgoing than incoming transition percentages, while others, such as end-stage conditions,  display the reverse pattern, indicative of directional flow in the population-level progression structure. \\ 
To identify key structural nodes, we quantified disease importance using degree and betweenness centrality. The highest-ranked diseases are summarised in Table~\ref{tab:top3_centrality}, with complete metrics reported in Supplementary Table~\ref{ST:centrality}.
End-stage renal disease exhibits the highest degree and betweenness centrality, identifying it as the most connected node and the principal bridge between distinct regions of the network despite its relatively low cohort prevalence. Cerebrovascular disease and acute kidney injury share the second-highest degree centrality; cerebrovascular disease also ranks second in betweenness centrality. Ventricular tachycardia ranks third in degree centrality but substantially lower in betweenness centrality, indicating a local hub role within a subset of the network rather than a global bridging function.

\begin{table}[h]
\centering
\small
\begin{tabular}{ccc}
\hline
\textbf{Rank} &\textbf{Degree} & \textbf{Betweenness} \\
\hline
1 & End Stage Renal Disease  & End Stage Renal Disease \\
2 & Cerebrovascular Disorders, Acute Kidney Injury & Cerebrovascular Disorders \\
3 & Ventricular Tachycardia & Acute Kidney Injury \\
\hline
\end{tabular}
\caption{Top 3 diseases ranked by betweenness and degree centrality in the estimated network. Details on centrality for other diseases are provided in Supplementary Table~\ref{ST:centrality}.}
\label{tab:top3_centrality}
\end{table}

\begin{figure}[h]
    \centering
    \includegraphics[width=1\textwidth]{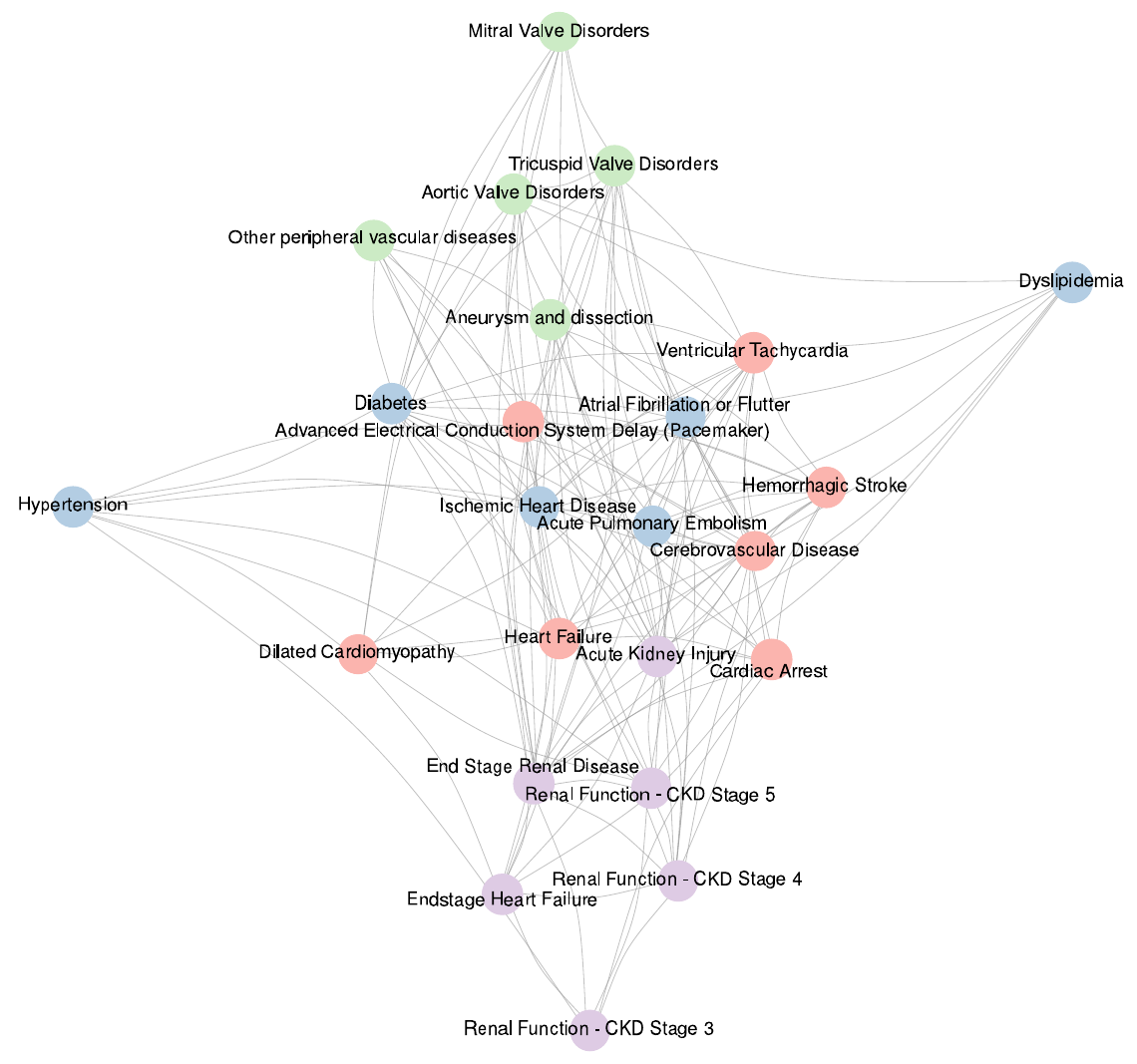}
        \caption{Community structure of the risk-adjusted cardiometabolic disease network. Node colours indicate community membership: \textit{light blue} = Cardiometabolic module; \textit{red} = Cardiac Rhythm and Conduction module; \textit{lilac} = Renal Disease module; \textit{light green} = Structural Cardiovascular module.}
    \label{fig:community_detection}
\end{figure}

The fast greedy modularity optimisation algorithm identified four distinct disease modules, shown in Figure~\ref{fig:community_detection}. The first module (Cardiometabolic, light blue) groups hypertension, diabetes, dyslipidaemia, and major cardiovascular events. The second module (Cardiac Rhythm and Conduction, red) groups conditions characterised by impaired myocardial contractility and conduction defects. The third module (Renal Disease, lilac) consolidates the chronic kidney disease progression pathway. The fourth module (Structural Cardiovascular, light green) groups structural and large-vessel pathologies.

Building on the detected community structure, we derived the individual-level normalised transition matrices (see Section~\ref{subsub:transition}). Averaging these across individuals yields a population-level transition matrix (Supplementary Figure~\ref{supp_mat:SF5}) that reveals strong dominance of within-community persistence in the cardiometabolic module. This module also acts as the primary destination for transitions originating from all other modules: the cardiac rhythm and conduction module and the renal disease module both exhibit higher outgoing transition probabilities toward the cardiometabolic domain than within their own communities. The structural cardiovascular module displays consistently low transition probabilities in all directions.



Clustering of individual-level normalised transition matrices identified four distinct progression dynamics; the corresponding mean transition matrices by cluster are displayed in Supplementary Figure~\ref{supp_mat:SF6} and characterised in Supplementary Table~\ref{ST:cluster_descriptive_transposed}. Cluster 1 is the largest group (n = 38915, 34.9\%) and is characterised by near-exclusive persistence within the cardiometabolic module, with mean transitions close to zero toward all other domains. Cluster 2 (n = 23321, 20.9\%) shows a more distributed pattern, with substantial bidirectional probability mass between the cardiometabolic and cardiac rhythm and conduction modules. Cluster 3 (n = 31062, 27.9\%) retains within-cardiometabolic persistence as the dominant pattern but exhibits a prominent secondary axis toward the cardiac rhythm and conduction module, with additional spread across the cardiovascular and renal disease domains. Cluster 4 is the smallest (n = 18125, 16.3\%) and is distinguished by a strong directional transition probability from the cardiometabolic module toward the renal disease module, alongside moderate within-cardiometabolic persistence.

\begin{figure}[h]
    \centering
    \includegraphics[width=0.85\textwidth]{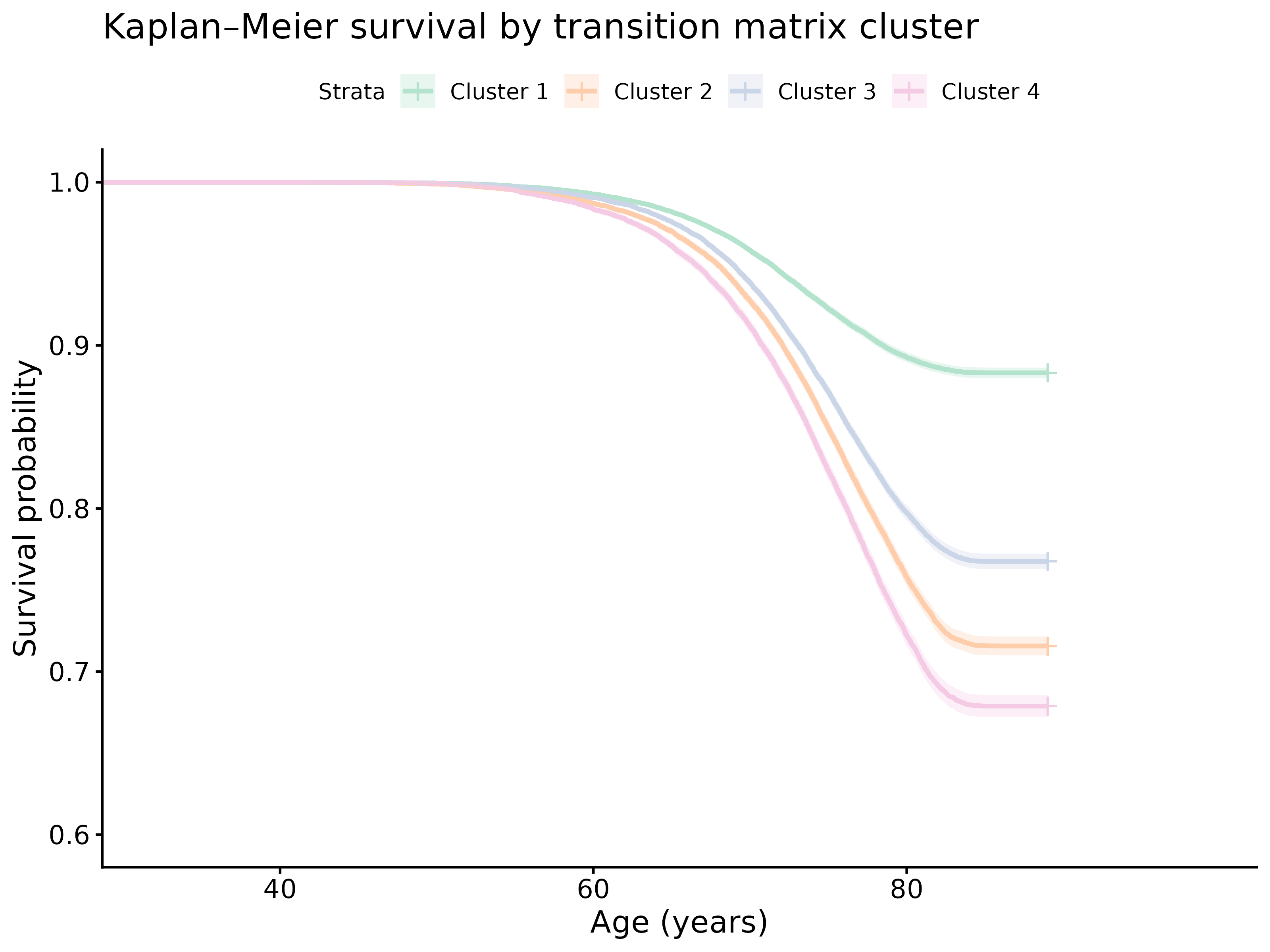}
    \caption{Kaplan-Meier survival curves stratified by transition cluster.}
    \label{Survival_by_cluster}
\end{figure}

Kaplan-Meier survival curves stratified by transition cluster revealed statistically significant differences in all-cause mortality across the four groups (Figure~\ref{Survival_by_cluster}). Curves diverge substantially after age 60, with cluster 1 exhibiting the most favourable survival trajectory and cluster 4 showing the worst survival outcome, aligning with its highest mortality rate (32.1\%) and greatest trajectory complexity. Clusters 2 and 3 occupy intermediate positions, with Cluster 2 showing slightly worse survival than Cluster 3 at older ages, consistent with the broader cross-community movement and higher disease burden observed in that group.

\section{Discussion}\label{sec:discussion}
Comorbidity networks provide a powerful framework for understanding relationships among diseases~\citep {Chiadi2023}. However, existing approaches face three critical limitations: the absence of temporal information, vulnerability to confounding by shared risk factors, and the inability to distinguish between direct and indirect associations. We developed a comprehensive methodological pipeline that addresses these limitations, demonstrating both methodological validity and clinical utility when applied to UK Biobank data comprising 24 cardiometabolic diseases and 76 risk factors. 

Our approach adapts Word2Vec embeddings to learn disease representations from patient event sequences, treating diseases as ``words" and patient sequences as ``sentences''. This embedding-based framework captures both semantic relationships from biomedical literature and temporal patterns from patient data, offering substantial advantages over aggregate prevalence statistics. The continuous embedding space enables gradient-based optimisation, transfer learning from large biomedical corpora, and direct incorporation of risk factor patterns. We then introduced a principled confounding identification framework that compares similarity matrices from jointly embedded diseases and risk factors against a disease-only baseline. This procedure successfully identified disease-specific confounder sets, as evidenced by the systematic link between changes in similarity and shared risk factor burden, demonstrating that classical co-occurrence statistics substantially overestimate direct pathophysiological similarity. These validated prior clinical knowledge, offer an interpretable, data-driven characterisation of the shared risk-factor structure underlying each disease pair. Finally, we employed a Gaussian Graphical Model with Lasso regularisation to estimate sparse networks capturing only direct conditional dependencies. By conditioning each disease's neighbourhood regression on its identified confounders, the GGM explicitly adjusts for shared risk-factor burden while preserving statistical power, limiting the regression to only relevant variables. The integration of embedding-derived prior knowledge into graphical models represents a methodological novelty and principled alternative to both arbitrary thresholding and undifferentiated covariate adjustment. The resulting sparse network revealed clinically meaningful disease relationships, with topological analysis identifying key pathological hubs and four distinct disease communities that align with established cardiometabolic taxonomy. 

Beyond methodological contributions, the proposed framework supports clinically relevant downstream analyses with direct implications for evidence-based multimorbidity management. At the network level, identifying diseases with high centrality provides a principled, data-driven basis for prioritising interventions on conditions whose effects can propagate across multiple pathophysiological domains. At the individual level, clustering patients by community transition patterns yielded four clusters with significantly divergent survival outcomes, demonstrating that these patient-specific representations carry prognostic information beyond standard clinical risk stratification. By simultaneously capturing multiple-disease progression patterns and their interplay, information that single-disease stratification would miss, this approach creates new opportunities for personalised risk assessment, supporting earlier detection and intervention for high-risk multimorbidity phenotypes~\cite{liang2024comorbidity}.
  
Our study faces several limitations that suggest directions for future development. From a methodological perspective, our method could benefit from a simulation study to formally assess the edge recovery performance of the estimated network, which is not currently provided. This reflects the inherent difficulty of generating realistic synthetic benchmarks for this setting: a valid simulation would need to encode a known disease association structure while simultaneously preserving the temporal ordering, co-occurrence patterns, and confounding structure characteristic of real longitudinal clinical sequences, a level of joint complexity beyond what standard network-simulation frameworks are designed to handle. 
We therefore instead provide empirical support for the validity of our approach at two levels: internally, through the statistically significant association between similarity changes and shared risk factor burden, and the robustness of this finding across threshold specifications; and externally, through the clinical interpretability of the estimated network and communities, and the prognostic signal carried by the downstream trajectory phenotypes.
In addition, although our framework captures temporal ordering, it does not explicitly model time intervals or causal directionality between events; extending the methodology with causal inference or temporally evolving network approaches could improve the characterisation of disease progression dynamics~\citep{pearl2009causality, prosperi2020causal, xu2019patient, kong_causal, dervic2025comorbidity, gardinazzi2025characterization}.
From an application perspective, residual confounding from unmeasured variables may persist despite adjustment for 76 risk factors, and future work could further enrich the framework by integrating genetic, proteomic, and metabolomic information~\citep{sibilio2025integrating, nam2020network}. Moreover, the embedding approach assumes that diseases occurring in similar temporal contexts share meaningful relationships, an assumption that may be less reliable for rare or clinically heterogeneous conditions. Finally, the downstream patient-level transition representation depends on the stability of the detected community structure, and alternative detection algorithms could be explored based on the analytical context and clinical question.

This work demonstrates that integrating longitudinal patient data with explicit risk-factor adjustment and advanced representation learning supports reliable, clinically useful analyses of comorbidity networks. By addressing key limitations in existing approaches, our framework provides a statistically rigorous foundation for characterising multimorbidity patterns, generalisable beyond cardiometabolic conditions to other domains. As electronic health records and population-scale longitudinal cohorts become increasingly available, methods for extracting structured knowledge from complex observational data will be essential for translating data into actionable clinical evidence, a direction to which this work contributes a replicable methodological foundation.




\section*{Acknowledgments}
The present research has been supported by MUR, grant Dipartimento di Eccellenza 2023-2027.
F. Ieva acknowledges the National Plan for NRRP Complementary Investments "Advanced Technologies for Human-centred Medicine" (PNC0000003). We also thank Carlo Andrea Pivato and Laura Savaré for their support in the phenotypes definitions. This study used data from the UK Biobank Resource under Application Number 102297.

\section*{Data and code availability}
UK Biobank data are accessible for health-related research in the public interest. 
The full analysis pipeline, implemented in Python and R, is publicly available at \url{https://github.com/AlessiaMapelli/Com_network_NLP}, together with simulated data enabling reproduction of all methodological steps and figures.

\newpage
\section*{Supplementary Material}
\setcounter{table}{0}          
\renewcommand{\thetable}{S\arabic{table}}

\setcounter{subsection}{0}
\renewcommand{\thesubsection}{S\arabic{subsection}}

\subsection*{Supplementary Tables}
\begin{table}[h]
\centering
\caption{Cardiometabolic phenotypes definition and MeSH mapping along with lifecourse incidence.}
\label{ST:table1}
\vspace{2mm}
\footnotesize The tables are available online at: \url{https://github.com/AlessiaMapelli/Com_network_NLP/blob/main/Supp_tabels/mapping_diseases.xlsx}
\end{table}

\begin{table}[h]
\centering
\caption{Risk factors phenotypes definition and MeSH mapping.}
\label{ST:table2}
\vspace{2mm}
\footnotesize The tables are available online at: \url{https://github.com/AlessiaMapelli/Com_network_NLP/blob/main/Supp_tabels/mapping_diseases.xlsx}
\end{table}

\begin{table}[ht]
\centering
\caption{Shared risk factors for each disease pair across different sensitivity thresholds ($\tau$).}
\label{ST:shared_rf_sensitivity}
\small
\begin{tabular}{|l|c|c|c|c|}
\hline
\textbf{Disease Pair} & \textbf{0.6} & \textbf{0.7} & \textbf{0.72*} & \textbf{0.8} \\
\hline
TVD $\leftrightarrow$ HTN  & 0.98 & 0.84 & \textbf{0.80} & 0.30 \\
TVD $\leftrightarrow$ DM   & 0.97 & 0.95 & \textbf{0.89} & 0.50 \\
MVD $\leftrightarrow$ HTN  & 0.99 & 0.94 & \textbf{0.84} & 0.50 \\
AKI $\leftrightarrow$ HTN  & 0.98 & 0.94 & \textbf{0.89} & 0.50 \\
MVD $\leftrightarrow$ DM   & 0.98 & 0.93 & \textbf{0.91} & 0.61 \\
AKI $\leftrightarrow$ DM   & 0.97 & 0.96 & \textbf{0.91} & 0.83 \\
CKD3 $\leftrightarrow$ MVD & 0.98 & 0.97 & \textbf{0.98} & 0.72 \\
CKD4 $\leftrightarrow$ MVD & 0.99 & 0.96 & \textbf{0.97} & 0.72 \\
CKD3 $\leftrightarrow$ TVD & 0.98 & 0.98 & \textbf{1.00} & 0.70 \\
MVD $\leftrightarrow$ CKD5 & 0.99 & 0.97 & \textbf{0.97} & 0.72 \\
IHD $\leftrightarrow$ AKI  & 1.00 & 1.00 & \textbf{0.99} & 0.90 \\
HTN $\leftrightarrow$ VT   & 0.95 & 0.80 & \textbf{0.76} & 0.43 \\
PVD $\leftrightarrow$ TVD  & 0.99 & 0.98 & \textbf{1.00} & 0.80 \\
HTN $\leftrightarrow$ AVD  & 0.99 & 0.97 & \textbf{0.92} & 0.52 \\
\hline
\textbf{Mean (SD)} & 0.98 (0.01) & 0.94 (0.06) & \textbf{0.92 (0.08)} & 0.63 (0.17) \\
\textbf{Wilcoxon p-value} & $7.66 \times 10^{-6}$ & $7.22 \times 10^{-4}$ & $\mathbf{3.83 \times 10^{-3}}$ & $6.20 \times 10^{-5}$ \\
\hline
\end{tabular}
\vspace{0.8em}
\footnotesize
\parbox{0.97\linewidth}{
\textit{Note:} Abbreviations: TVD = Tricuspid Valve Disorders, MVD = Mitral Valve Disorders, DM = Diabetes, HTN = Hypertension, AKI = Acute Kidney Injury, VT = Ventricular Tachycardia, AVD = Aortic Valve Disorders, CKD3-5 = Renal Function - CKD 3-5, IHD = Ischemic Heart Disease, PVD = Other Peripheral Vascular Diseases. *Main analysis threshold. Values indicate the proportion of shared risk factors retained at each threshold. Wilcoxon rank-sum p-values compare the first 14 highly affected disease pairs against all remaining pairs at each threshold.
}
\end{table}

\newpage

\begin{table}[h]
\centering
\begin{tabular}{lcc}
\hline
\textbf{Disease} & \textbf{Degree Centrality Rank} & \textbf{Betweenness Centrality Rank} \\
\hline
End stage Renal Disease ,  & 1 & 1 \\
Cerebrovascular Disease & 2 & 2 \\
Acute Kidney Injury & 2 & 3 \\
Ventricular Tachycardia & 3 & 7 \\
Diabetes & 4 & 4 \\
Ischemic Heart Disease & 5 & 5 \\
Atrial Fibrillation or Flutter & 5 & 6 \\
Acute Pulmonary Embolism & 5 & 12 \\
Tricuspid Valve Disorders & 5 & 10 \\
AECSD (Pacemaker)  & 6 & 8 \\
Heart Failure & 7 & 9 \\
Aneurysm and Dissection& 7 & 11 \\
Hemorrhagic Stroke & 7 & 20 \\
Renal Function - CKD 5 & 8 & 13 \\
Cardiac Arrest & 8 & 14 \\
Renal Function - CKD 4 & 8 & 17 \\
Aortic Valve Disorders & 8 & 18 \\
Dilated Cardiomyopathy  & 9 & 16 \\
End stage Heart Failure  & 9 & 15 \\
Hypertension & 10 & 19 \\
Mitral Valve Disorders & 10 & 23 \\
Dyslipidemia & 11 & 24 \\
Other Peripheral Vascular Diseases & 11 & 22 \\
Renal Function - CKD 3 & 12 & 21 \\
\hline
\end{tabular}
\caption{Disease rankings based on degree centrality and betweenness centrality. AECSD= Advanced Electrical Conduction System Delay}
\label{ST:centrality}
\end{table}

\newpage
\begin{table}[h]
\centering
\small
\begin{tabular}{lcccc}
\hline
\textbf{Metric} & \textbf{Cluster 1} & \textbf{Cluster 2} & \textbf{Cluster 3} & \textbf{Cluster 4} \\
\hline
N (\%) & 38,915 (34.9) & 23,321 (20.9) & 31,062 (27.9) & 18,125 (16.3) \\
Dead (\%) & 11.7 & 28.4 & 23.2 & 32.1 \\
Age at death (yrs) & 71.6 (6.7) & 73.5 (6.6) & 73.2 (6.3) & 73.1 (6.7) \\
Distinct diseases & 3.33 (0.64) & 5.13 (2.21) & 4.72 (1.61) & 4.38 (1.77) \\
Total events & 3.61 (1.32) & 6.19 (3.72) & 5.27 (2.53) & 4.83 (2.54) \\
Entropy & 0.02 (0.09) & 0.63 (0.18) & 0.50 (0.14) & 0.56 (0.16) \\
\hline
\end{tabular}
\caption{Descriptive statistics for each trajectory-based patient cluster. Dead (\%) reports the proportion of patients who died during follow-up. Age at death, number of unique diseases, total number of diseases, and normalised transition entropy are reported as mean (sd) standard deviation. Normalised entropy quantifies the diversity of community transitions for each patient, with values close to 0 indicating near-exclusive persistence within a single community and values close to 1 indicating uniformly distributed transitions across all communities.}
\label{ST:cluster_descriptive_transposed}
\end{table}

\setcounter{figure}{0}          
\renewcommand{\thefigure}{S\arabic{figure}}

\clearpage

\subsection*{Supplementary Figure}
\begin{figure}[H]
    \centering
    \includegraphics[width=1\textwidth]{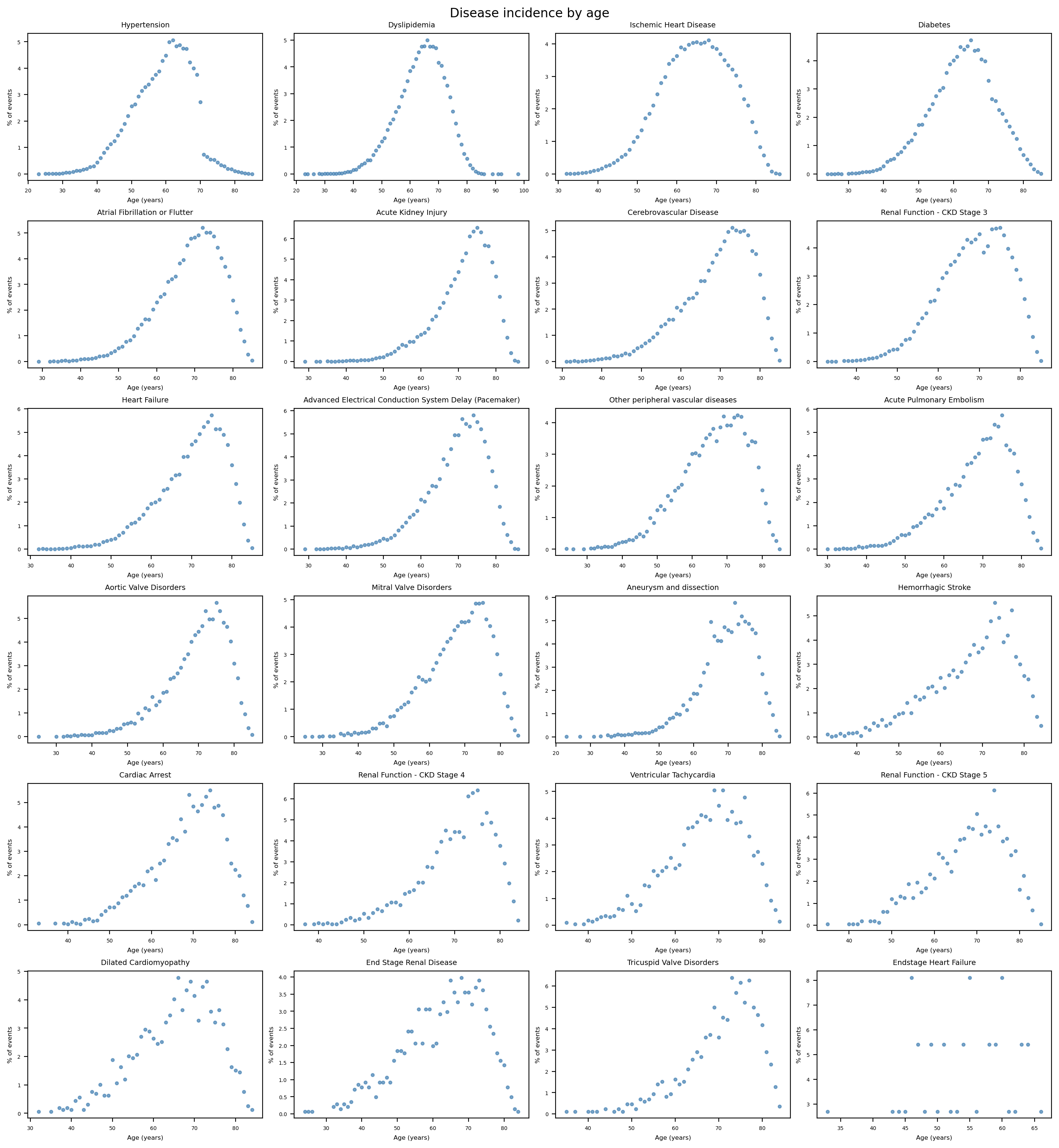}
    \caption{Age distribution of disease incidence across 24 cardiometabolic diseases. Each panel shows the number of incident cases by age (in years) for a specific disease. The y-axis scale varies across panels to accommodate different incidence rates.}
    \label{supp_mat:SF0}
\end{figure}

\newpage

\begin{figure}[h]
    \centering
    \includegraphics[width=0.9\textwidth]{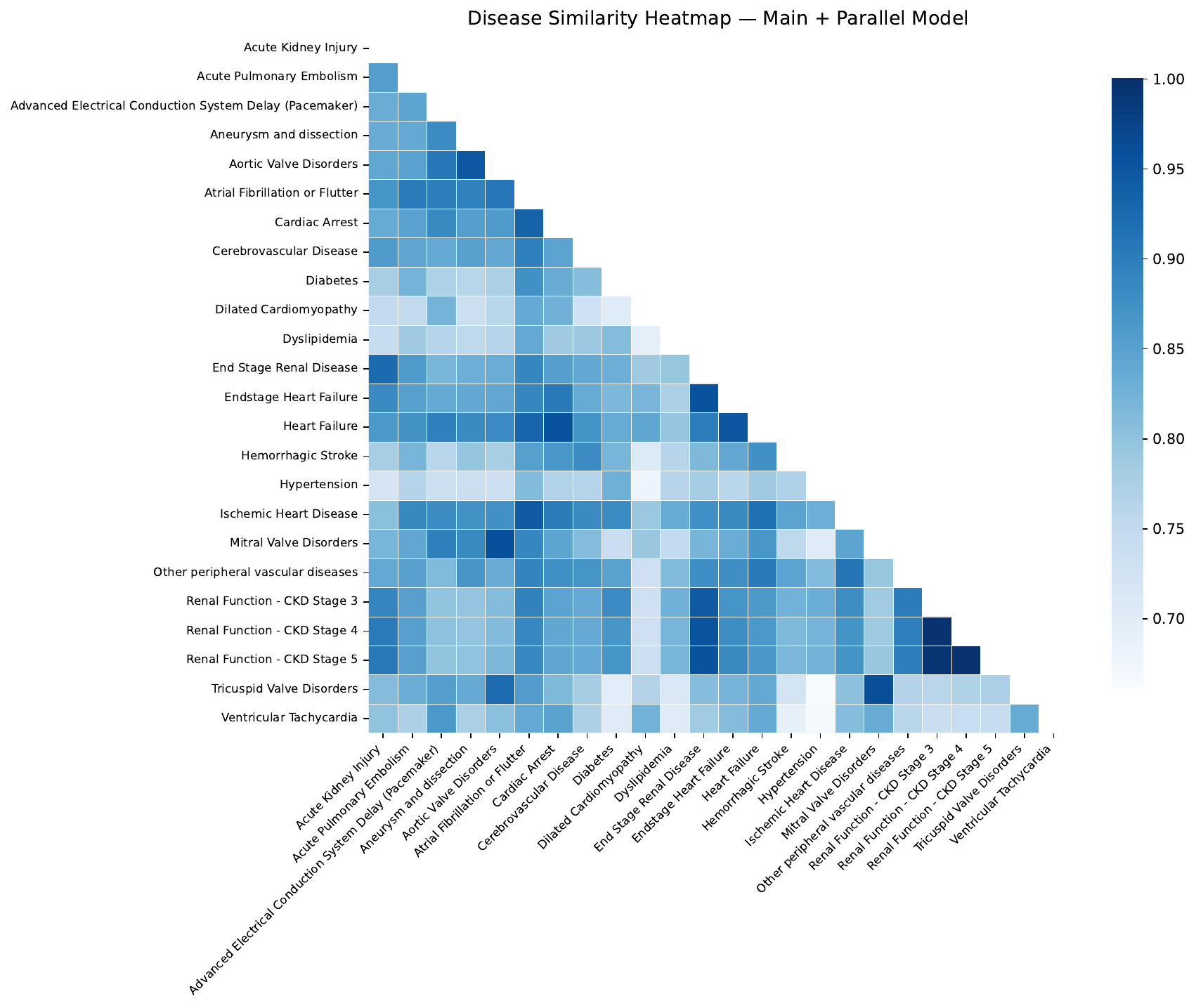}
    \caption{Cosine similarity matrix for 24 cardiometabolic diseases computed from primary sequence embeddings. Darker colors indicate stronger similarities.}
    \label{supp_mat:SF2}
\end{figure}

\begin{figure}[H]
    \centering
    \includegraphics[width=1\textwidth]{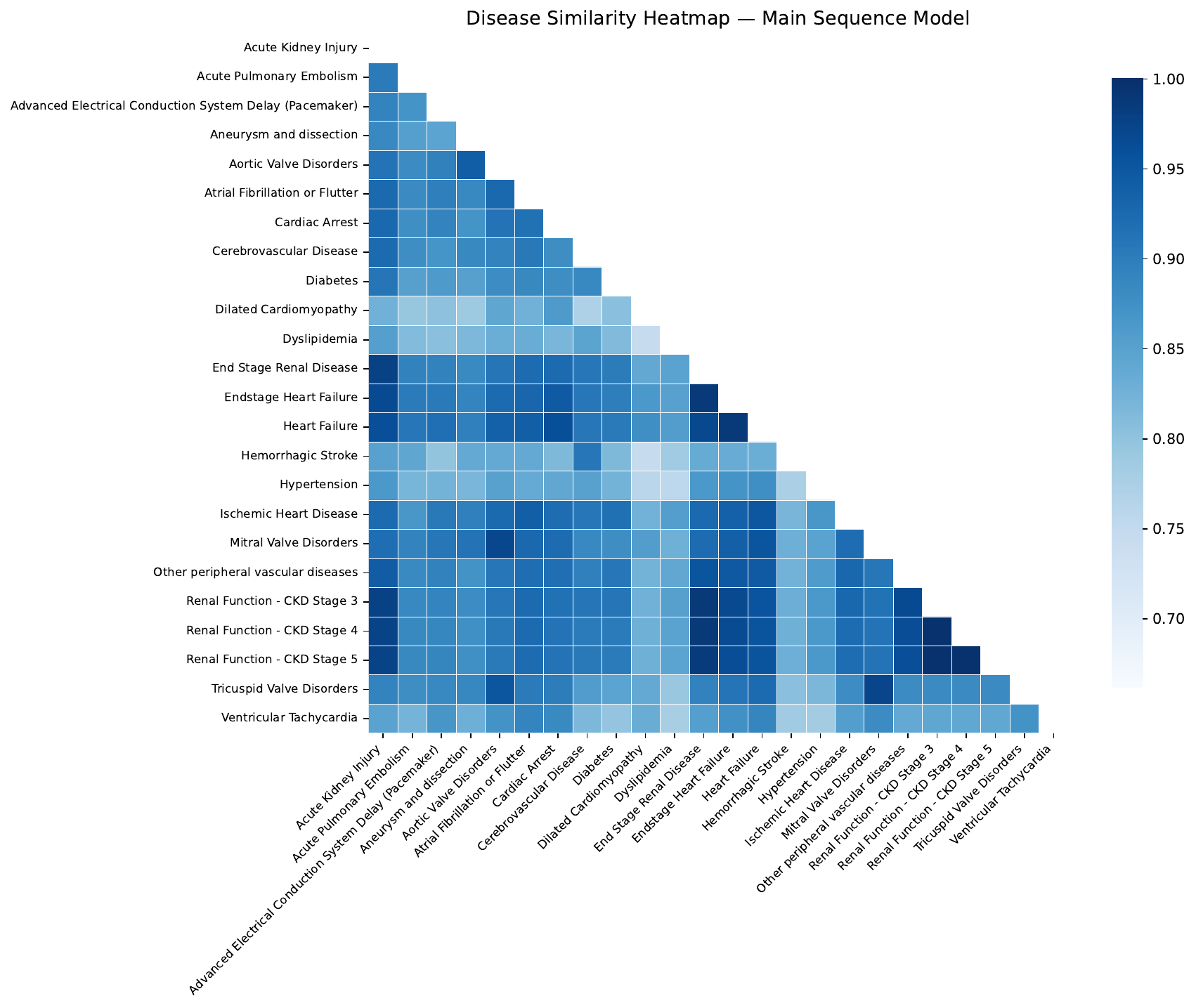}
    \caption{Cosine similarity matrix for 24 cardiometabolic diseases computed from secondary sequence embeddings. Darker colors indicate stronger similarities.}
    \label{supp_mat:SF1}
\end{figure}

\newpage

\begin{figure}[H]
    \centering
\includegraphics[angle=270,width=0.6\textwidth]{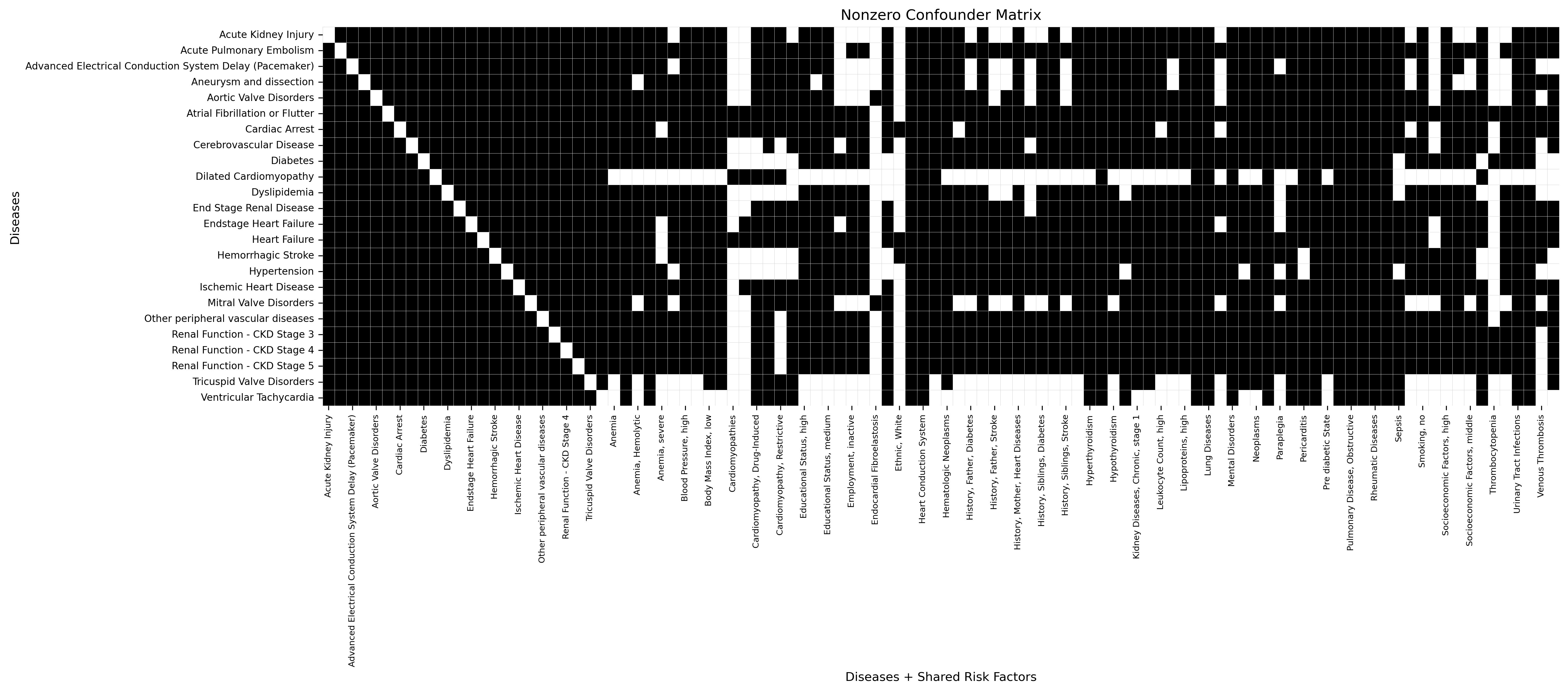}
    \caption{Disease-specific confounder sets used as prior clinical knowledge in the GGM neighbourhood regression. A filled (black) cell indicates that the risk factor was identified as a confounder for that disease and is included as a covariate in its neighbourhood regression.}
    \label{supp_mat:SF3}
\end{figure}

\newpage
\begin{figure}[H]
    \centering
    \includegraphics[width=1\textwidth]{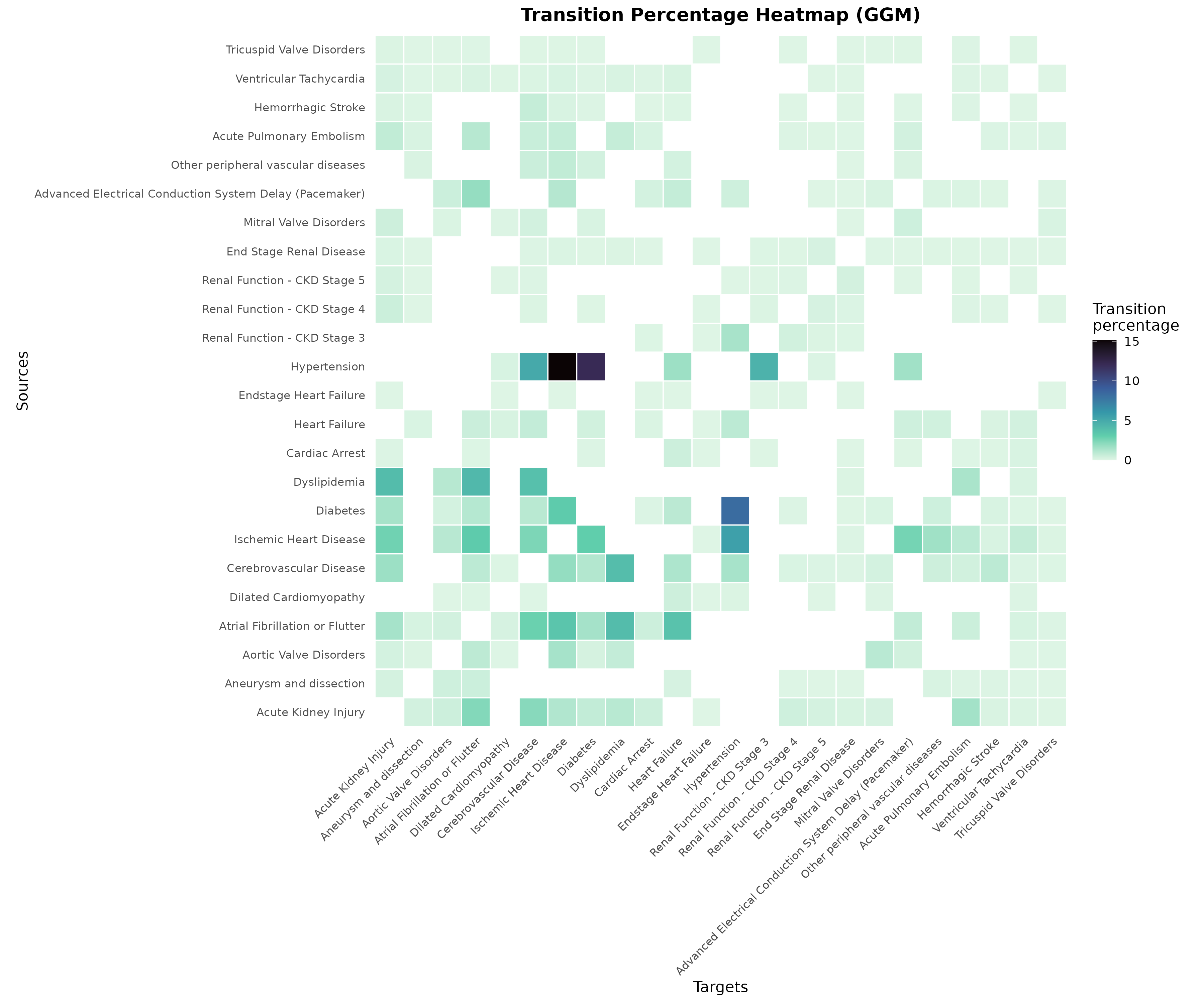}
    \caption{Disease-level transition matrix across 24 cardiometabolic diseases. Darker colours indicate higher transition percentages, masked by the edges found in the estimated network.}
    \label{supp_mat:SF4}
\end{figure}

\newpage
\begin{figure}[h]
    \centering
    \includegraphics[width=0.7\textwidth]{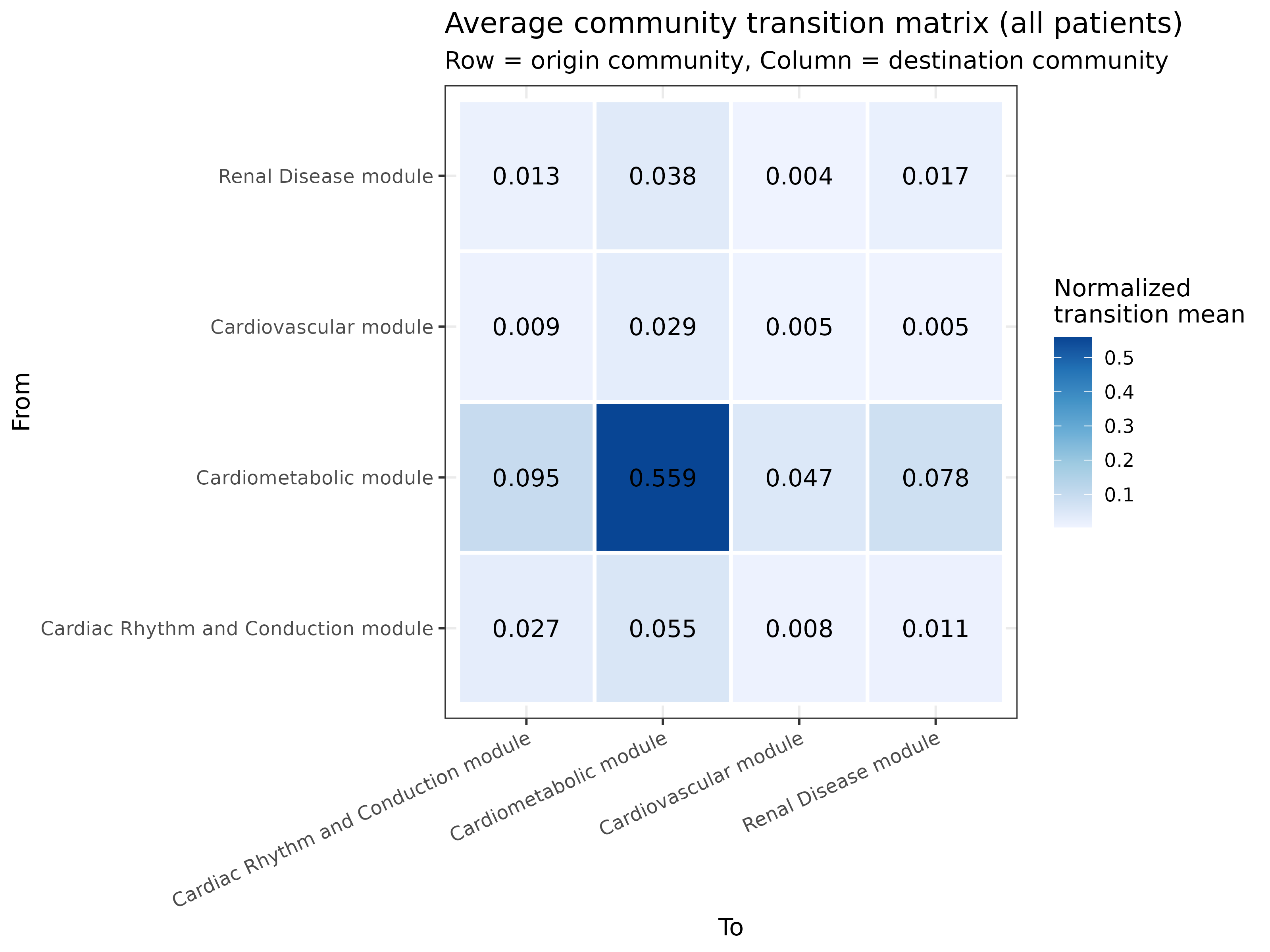}
    \caption{Population-level average community transition matrix. Darker colours indicate higher transition mean.}
    \label{supp_mat:SF5}
\end{figure}

\begin{figure}[h]
    \centering
    \includegraphics[width=\textwidth]{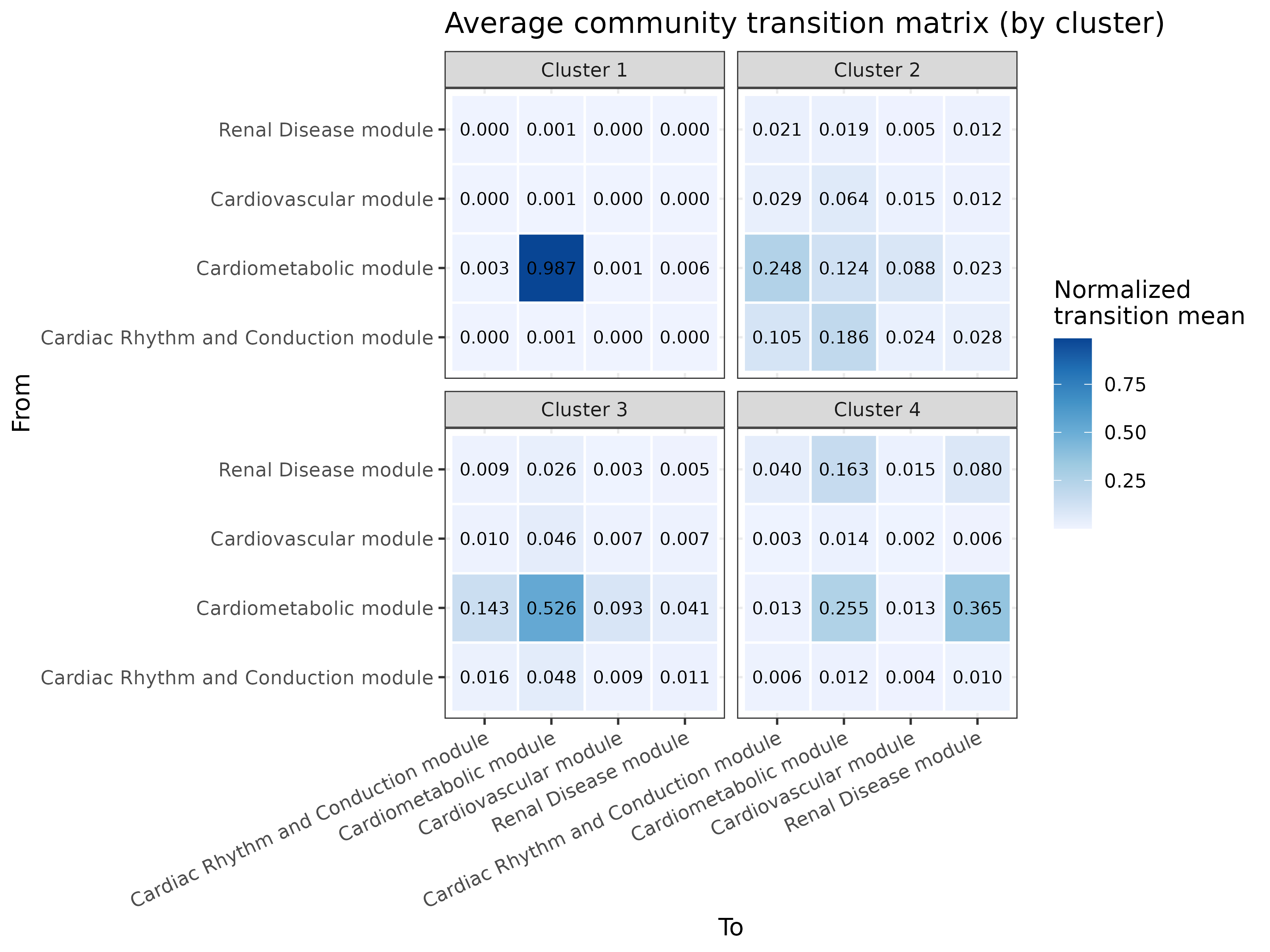}
    \caption{Mean community transition matrices for each of the four trajectory-based patient clusters. Each heatmap displays the average transition probability matrix computed over patients assigned to that cluster.}
    \label{supp_mat:SF6}
\end{figure}

\clearpage

\bibliographystyle{biorefs}
\bibliography{bibliography}

\end{document}